\documentclass[twocolumn, pra, reprint, nofootinbib, superscriptaddress, showkeys, showpacs, amsmath, amssymb]{revtex4-2}

\usepackage{graphicx,color}
\usepackage{bm}
\usepackage{graphicx}
\usepackage{dcolumn}
\usepackage{bm}
\usepackage{color}
\usepackage{soul}
\usepackage{tikz}
\usepackage[bookmarksnumbered=true]{hyperref} 
\hypersetup{colorlinks = true,linkcolor = blue,anchorcolor = blue,citecolor = blue, filecolor = blue,urlcolor = blue}
\usetikzlibrary{shapes,arrows,positioning,fit,backgrounds,calc}
\usetikzlibrary{shapes.geometric, arrows}
\usepackage[mathlines]{lineno}

\begin{document}

\preprint{APS/123-QED}

\title{Universal behavior of the  Covid-19 tails: Inverse power-law distribution}  

\author{E. Aydiner}
\email{ekrem.aydiner@istanbul.edu.tr} 

\affiliation{Theoretical and Computational Physics Research Laboratory, İstanbul University, 34134 İstanbul,  Türkiye}

\affiliation{Department of Physics, İstanbul University, 34134 İstanbul, Türkiye}

\author{E. Yılmaz} 
\email{erkany@ogr.iu.edu.tr} 

\affiliation{Theoretical and Computational Physics Research Laboratory, İstanbul University, 34134 İstanbul, Türkiye}

\affiliation{Institute of Graduate Studies in Science, Istanbul University, 34134 İstanbul, Türkiye}

\date{2 April 2024, İstanbul, Revised: 5 July 2024, Ankara}

\begin{abstract}
Power-law distribution is one of the most important laws known in nature. Such a special universal behavior is known to occur in very few physical systems. In this Letter, we analyzed the mortality distribution of the Covid-19 pandemic tails for different countries and continents to discuss the possible universal behavior of the pandemic. Surprisingly, we found that the mortality distribution of Covid-19 final i.e., the latest tails in 2023 follows inverse power-law decays. These universal behaviors for the pandemic are reported in the present work for the first time. Additionally, 
we showed that these mortality tails also decay with time obeying to the inverse power-law.

\end{abstract}


\maketitle


\section{Introduction} \label{Introduction}


Power-law distribution is given in the form
\begin{eqnarray} \label{power_form}
	f (x) \sim x^{\pm \alpha}
\end{eqnarray}
where $\alpha$ is the exponent which depends on the medium and $x$ is the continuous or stochastic spatial or time variable. The negative sign for $\alpha$, Eq.(\ref{power_form}) is called the inverse power-law distribution.

As it is known, power-law distribution is a law that has universal characteristics. Power-law distributions occur in an extraordinarily diverse range of phenomena. This law has universality and exists in various fields, from physics to biology, from political sciences to economics. For example; income distribution \cite{Pareto1897}, 
population of city \cite{Auerbach_1913,Zipf_1949}, 
the sizes of earthquakes \cite{Gutenberg_1944}, 
the frequency of use of words in any human language \cite{Estoup_1916}, diameter of moon craters \cite{Neukum_1994},
the number of citations received by papers \cite{Price_1965}, 
the number of hits on web pages \cite{Adamic_2000}, intensity of solar flares \cite{Lu_1991}, intensity of wars \cite{Roberts_1998}, the number of “hits” received by web sites \cite{Adamic_2000}, the frequency of occurrence of personal names in most cultures \cite{Zanette_2001}, the numbers of species in biological taxa \cite{Yule_1922}, connection distribution of the airports and internet web sites \cite{Albert_2002} and so on.

It should be noted that power laws in physics also occur in many situations other than the statistical distributions of quantities. For instance, Kepler's law \cite{Kepler_1609}, Newton \cite{Newton_1687tb} law for gravity, Coulomb's famous law, and the Stefan–Boltzmann law \cite{Stefan_1879,Boltzmann_1884} has a power-law form. Similarly, power-law behavior can be seen in the second-order phase transition where correlation functions and other physical quantities near the critical points decay with power-law exponents. Supercritical state of matter and supercritical fluids where supercritical exponents of heat capacity and viscosity decay with power-law \cite{Bolmatov_2013}. In addition to the examples mentioned above, power-law behavior may occur for various self-organized critical systems \cite{Bak_1987} such as forest fires \cite{Bak_1990_forest,Drossel_1992} and avalanche phenomena \cite{Bak_1987}.

In this Letter, we reported that the inverse power-law distribution can be seen in Covid-19 tails. As known, Covid-19, which started in 2020, was declared a pandemic by the World Health Organization \cite{March2020}. Between 2020 and 2023, it can be said that it is recorded that approximately one billion people were infected and more than seven million people lost their lives \cite{March2023}. We examined the final i.e., the latest tails in 2023 of Covid-19, that is, a short region near the end of the Covid-19 pandemic in 2023, and we showed that the distribution function in this region complies with the inverse power-law. In addition, the time dependence of deaths occurring in this tail region will be examined in detail.

This work is organized as follows: In Section \ref{Power_law}, we will analyze the power-law behavior of the Covid-19 morality data tails for various countries and continents. In Section \ref{Decay_law}, we will analyze the time dependence of the mortality data of the Covid-19 tails for the same regions. Finally, we close with some discussion and conclusion in the last section.

\section{Power-law distribution of the Covid-19 tails} \label{Power_law}

We analyze eight countries and continental data for Covid-19 tails in the present work. Firstly, we consider Germany (DEU), Espany (ESP), 
\begin{figure}[h!]  
    \centering
    \includegraphics[width=4.2cm]{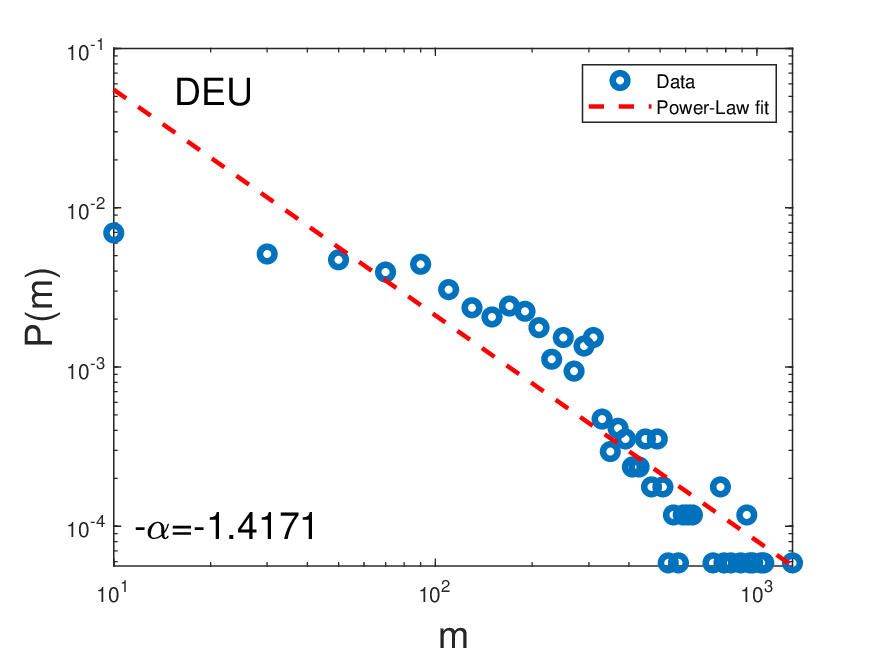}
    \includegraphics[width=4.2cm]{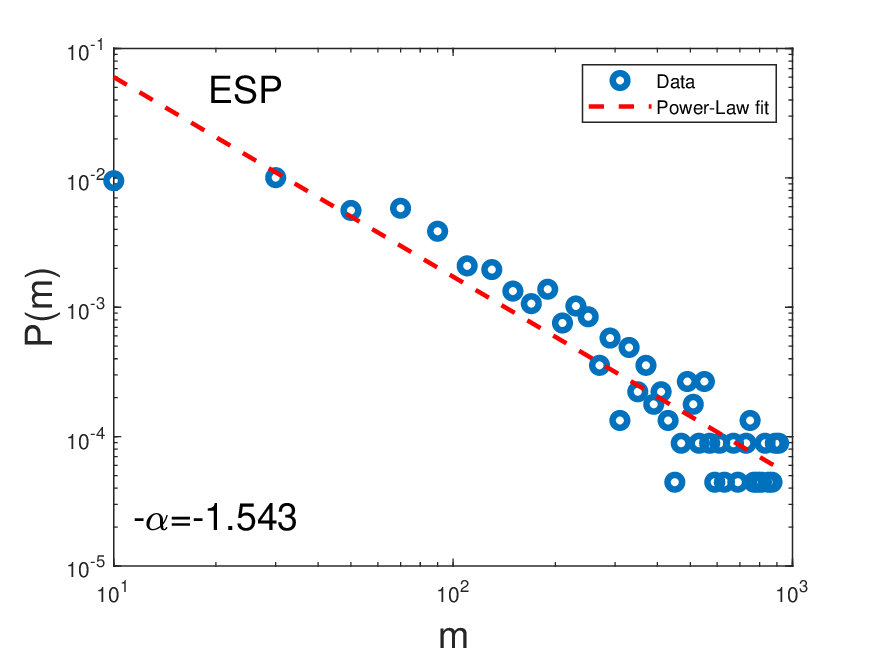}
    \includegraphics[width=4.2cm]{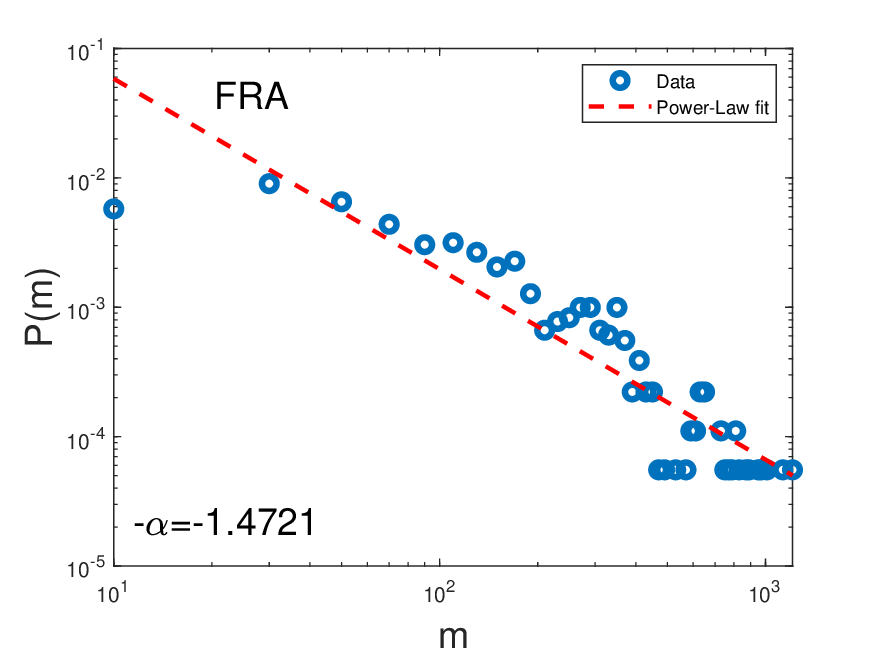}
    \includegraphics[width=4.2cm]{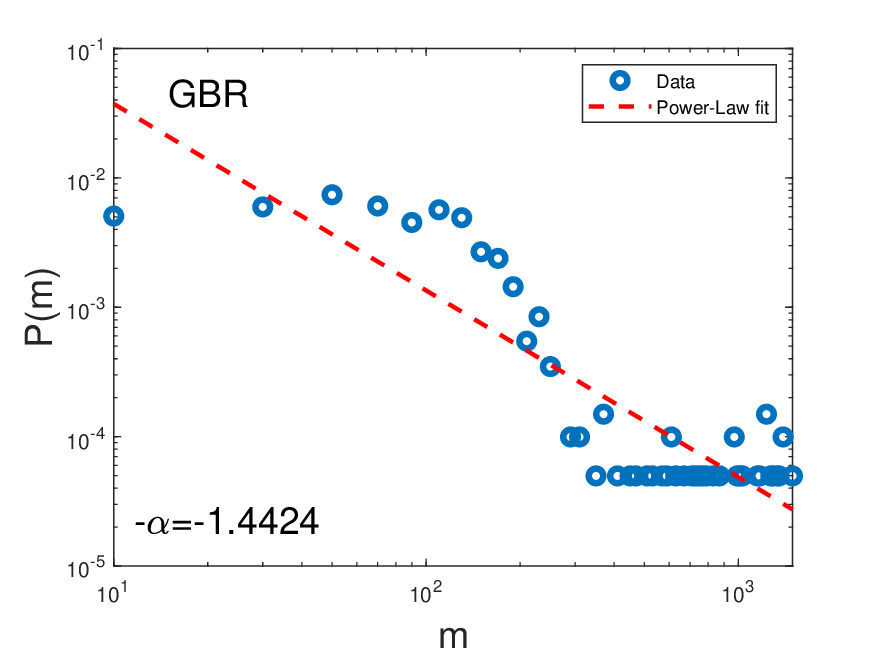}
    \includegraphics[width=4.2cm]{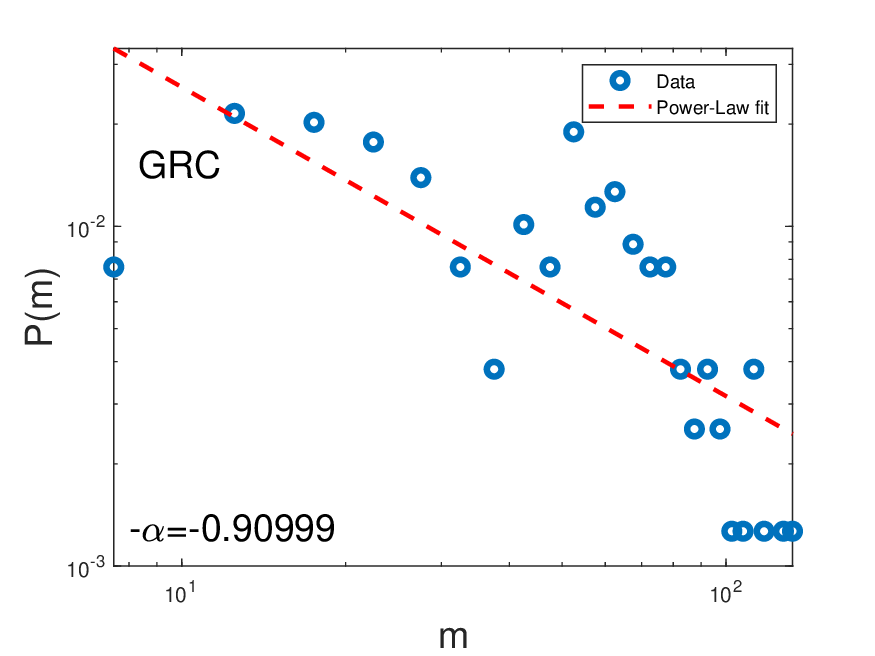}
    \includegraphics[width=4.2cm]{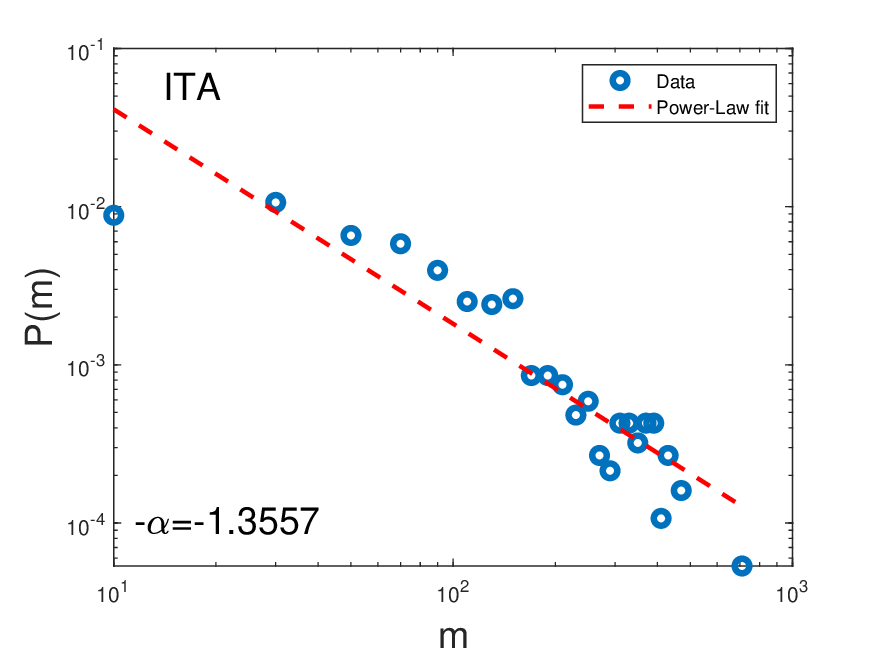}
    \includegraphics[width=4.2cm]{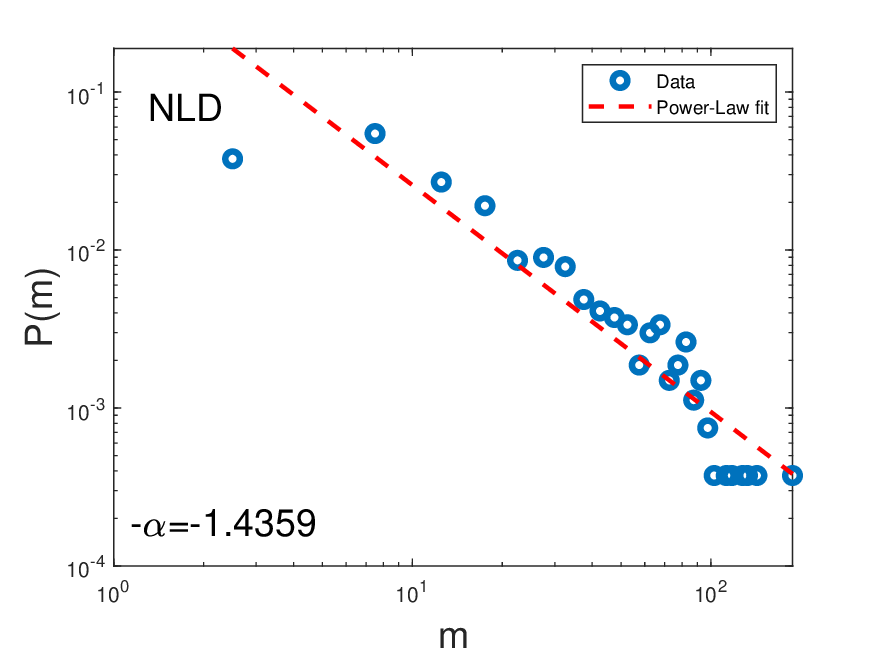}
    \includegraphics[width=4.2cm]{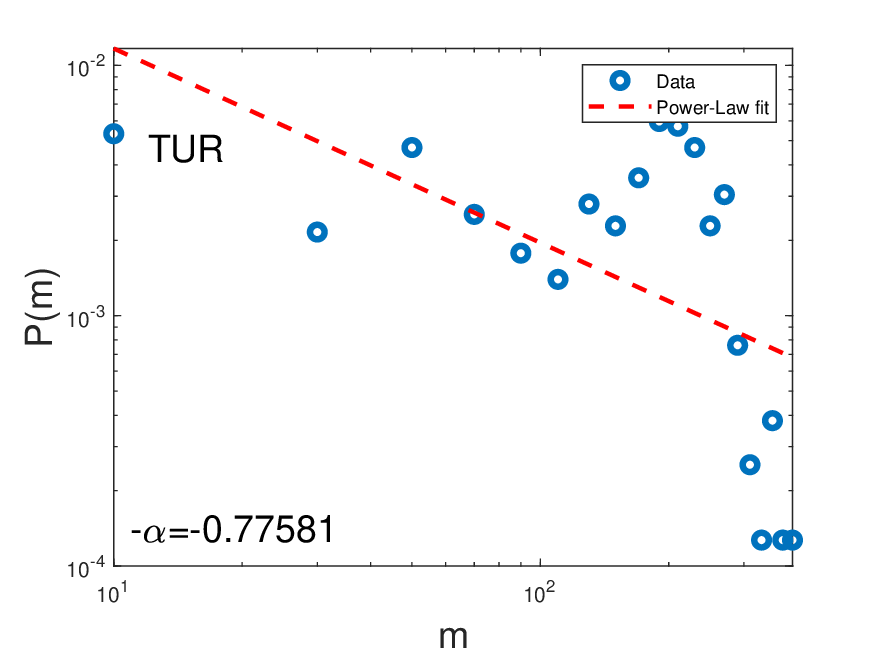}
    \caption{Mortality distribution $P(m)$ versus mortality frequency $m$ for eight countries in (a) DEU, in (b) ESP, in (c) FRA, in (d) GBR, in (e) GRC, in (f) ITA, in (g) NLD, and in (h) TUR. }  \label{Europe_Pm}
\end{figure}
France (FRA), United Kingdom (GBR), Greece (GRC), Italy (ITA), Holland (NLD), and Türkiye (TUR) mortality data for the period of the end of the pandemic \cite{March2023}. We plotted the data of these countries at $\log-\log$ scale in Fig.\ref{Europe_Pm}. As can be seen from the figure, the data of these examined countries comply with the power law given in Eq.(\ref{power_form}):
\begin{eqnarray} \label{inverse_power_m}
	P (m) \sim m^{ - \alpha} \quad \text{with} \quad 0 < \alpha <2
\end{eqnarray}
where $m$ denotes mortality frequency, and $\alpha$ is the exponent that characterizes mortality rates for each country. This universal behavior is known as the inverse power-law. 
\begin{figure} [h!] 
    \centering
    \includegraphics[width=4.2cm]{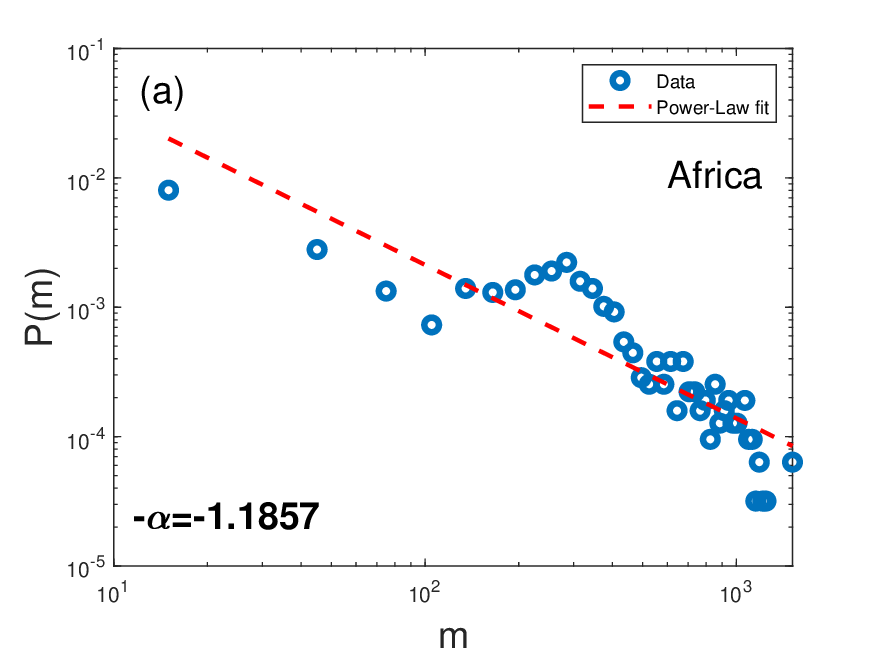}
    \includegraphics[width=4.2cm]{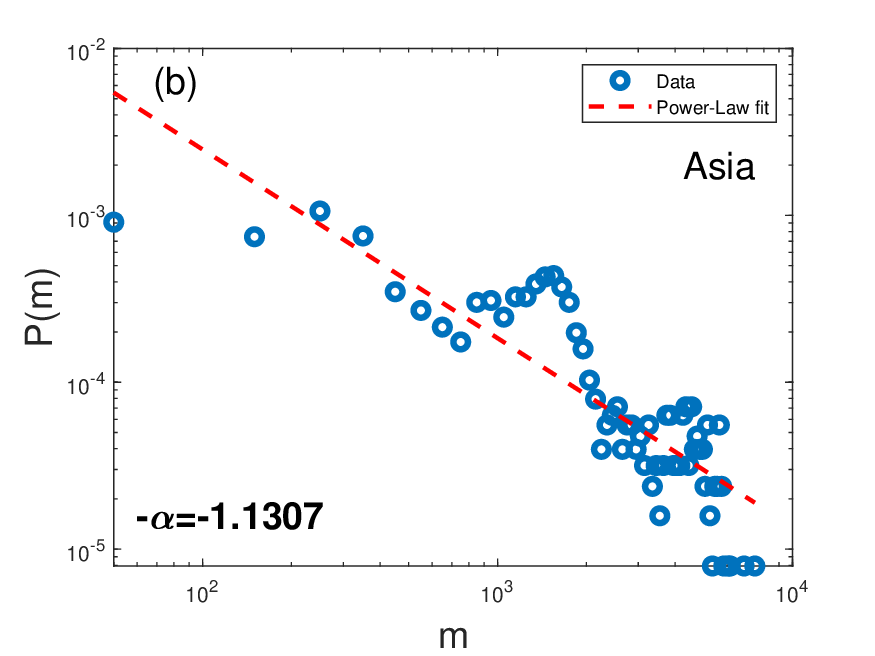}
    \includegraphics[width=4.2cm]{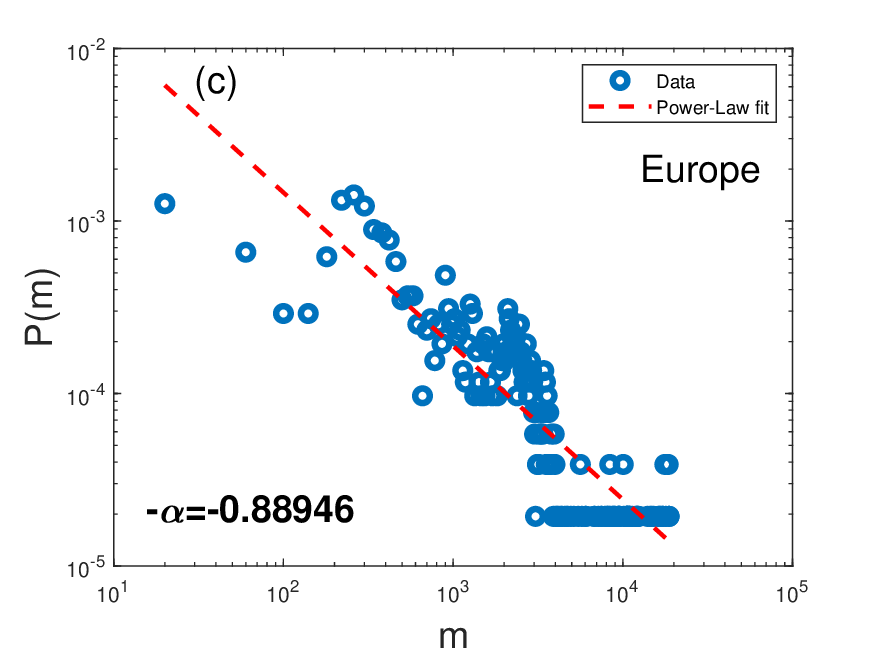}
    \includegraphics[width=4.2cm]{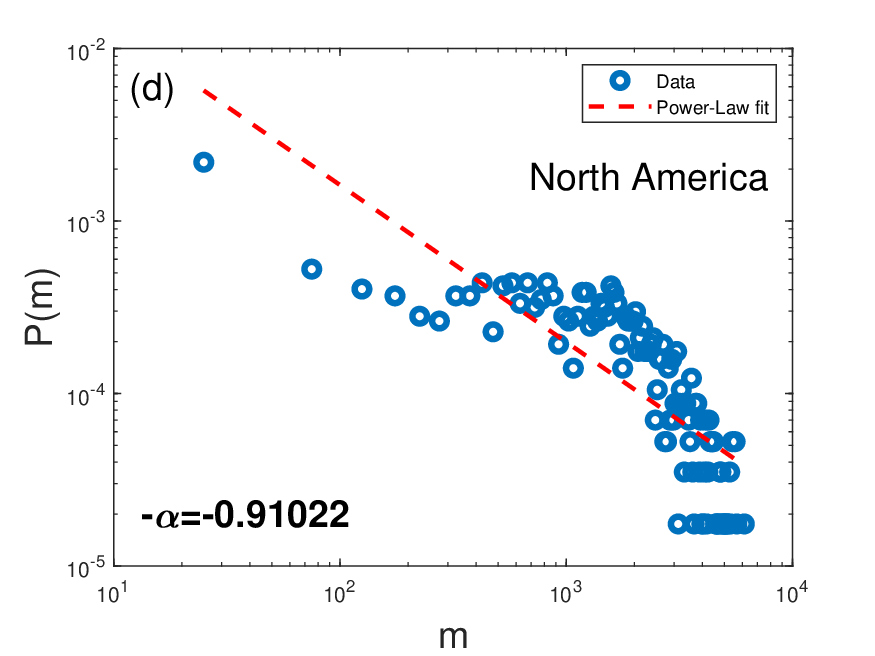}
    \includegraphics[width=4.2cm]{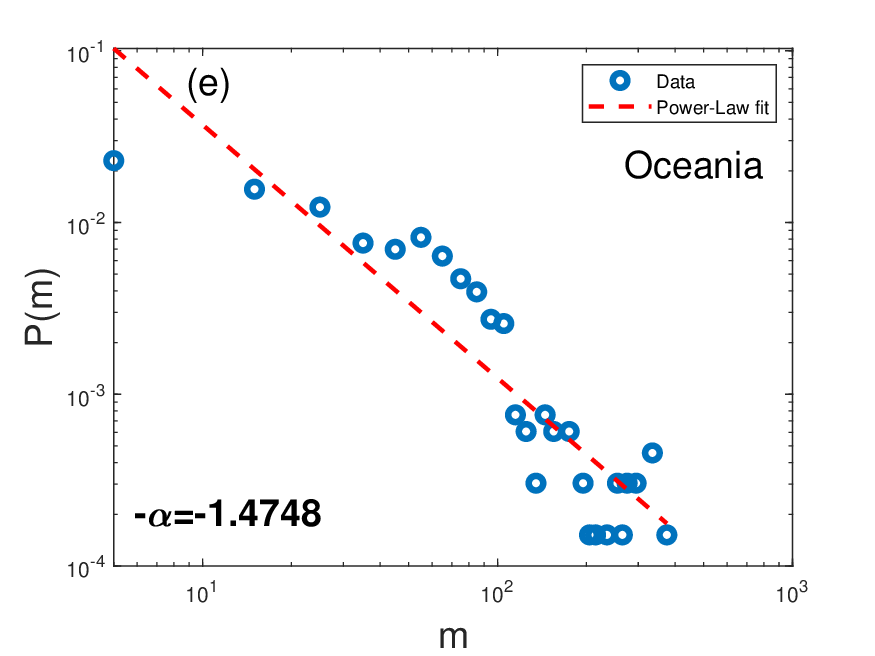}
    \includegraphics[width=4.2cm]{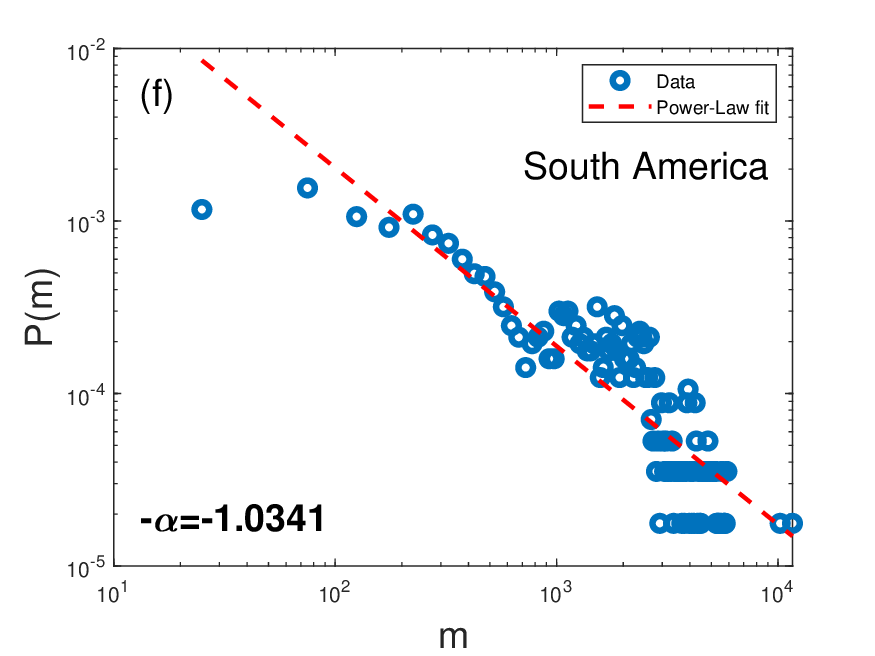}
    \caption{Mortality distribution $P(m)$ versus mortality frequency $m$ for six continentals in (a) Africa, in (b) Asia, in (c) Europe, in (d) North America, in (e) Oceania and in (f) South America. }  \label{contin_Pm}
\end{figure}
As can be seen from the sub-figures, $\alpha$ values take values between 0 and 2 for countries. For instance; for DEU $\alpha=1.4171$, for ESP  $\alpha=1.543$, for FRA $\alpha=1.4721$, for GBR $\alpha=1.4424$, for GRC $\alpha=0.90999$, for ITA $\alpha=1.3557$, for NLD $\alpha=1.4359$, and for TUR $\alpha=0.77581$.  

It is worth noting that the value of the exponents may not be related to the spread and death rate of Covid-19 for every country. The value of this parameter is valid only for the Covid-19 queue. We should state that this parameter represents the distribution of deaths in this tail region according to death frequency.

We also looked at the death distributions for Covid-19 outbreaks in six different continents to test whether the power-law distribution obtained for countries varies for wider geographies, in other words, whether this characteristic behavior is universal. It can be seen that the distribution functions of the continents given in Fig.\ref{contin_Pm} also comply with Eq.(\ref{inverse_power_m}). For continents, similarly, the $\alpha$ value takes values between 0 and 2. For instance: for Africa $\alpha=1.1857$, for Asia $\alpha=1.1307$, for Europe $\alpha=0.88946$, for North America $\alpha=0.91022$, for Oceania $\alpha=1.14748$, for South America $ \alpha=$1.0341.

Our results on the Covid-19 tail show that death distributions during the pandemic termination period have a universal character and obey the inverse power law, as in many interesting phenomena in nature \cite{Pareto1897,Auerbach_1913,Zipf_1949,Gutenberg_1944,Estoup_1916,Neukum_1994,Price_1965,Adamic_2000,Lu_1991,Roberts_1998,Adamic_2000,Zanette_2001,Yule_1922,Albert_2002}. This universal behavior was obtained for the first time in this study of the Covid-19 tail.

\section{Time-dependent Decay of the Covid-19 tails} \label{Decay_law}

As in the previous section, we analyze eight same countries and the same continental data to discuss the time dependence of the Covid-19 tails. Therefore, we also analyze the time dependence of the Covid-19 data tails for the same countries given in the previous section.  
\begin{figure}[h!]  
    \centering
    \includegraphics[width=4.2cm]{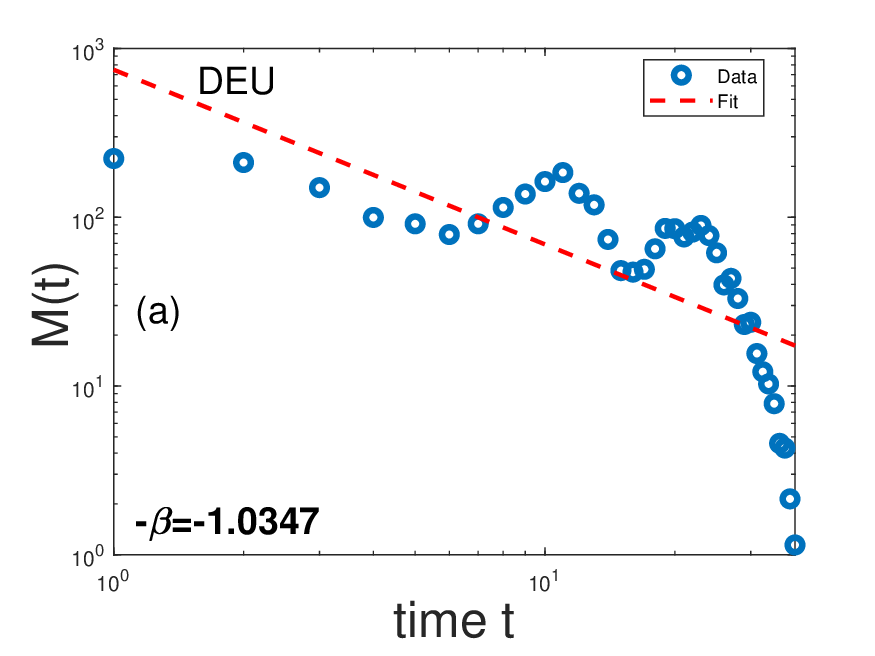}
    \includegraphics[width=4.2cm]{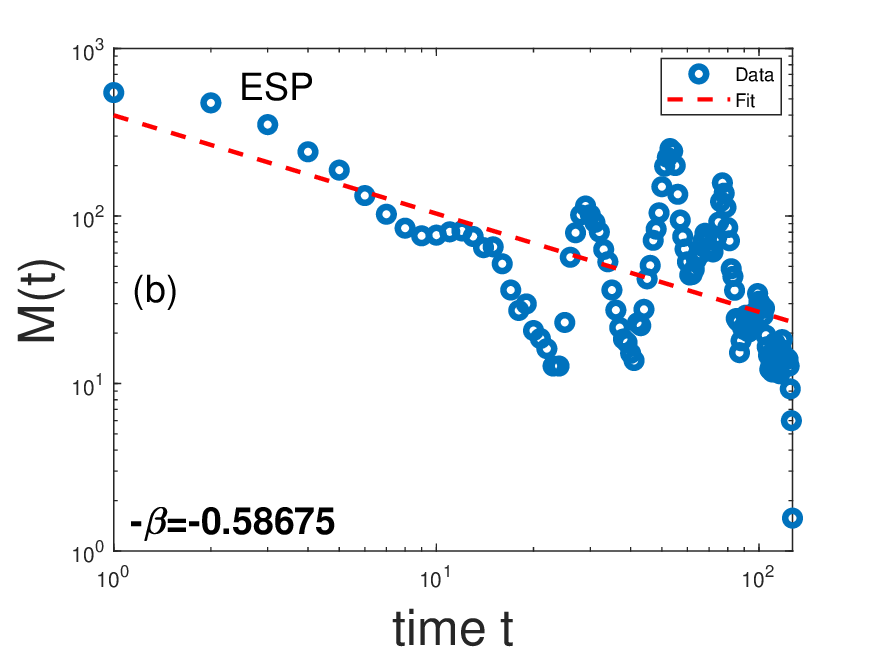}
    \includegraphics[width=4.2cm]{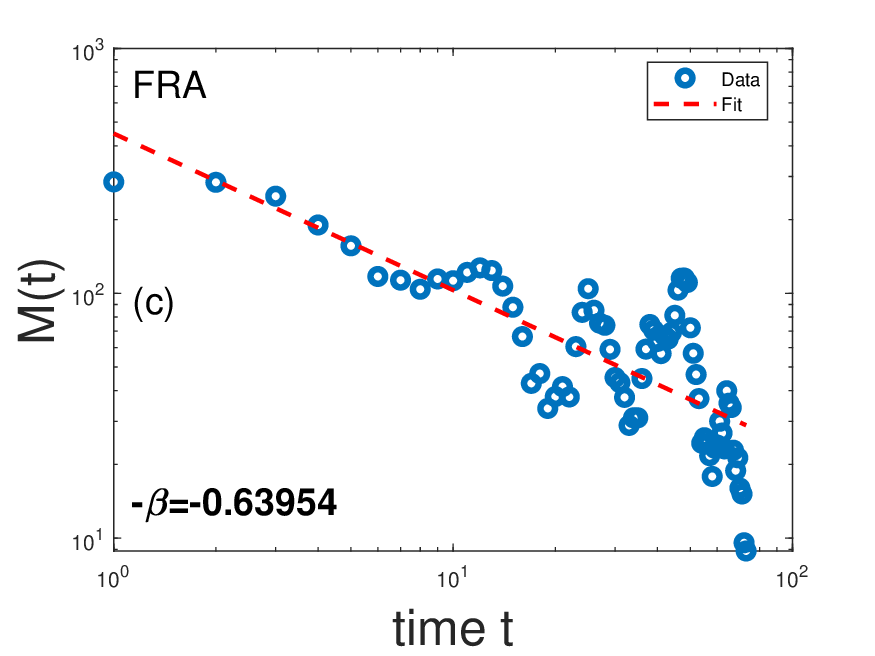}
    \includegraphics[width=4.2cm]{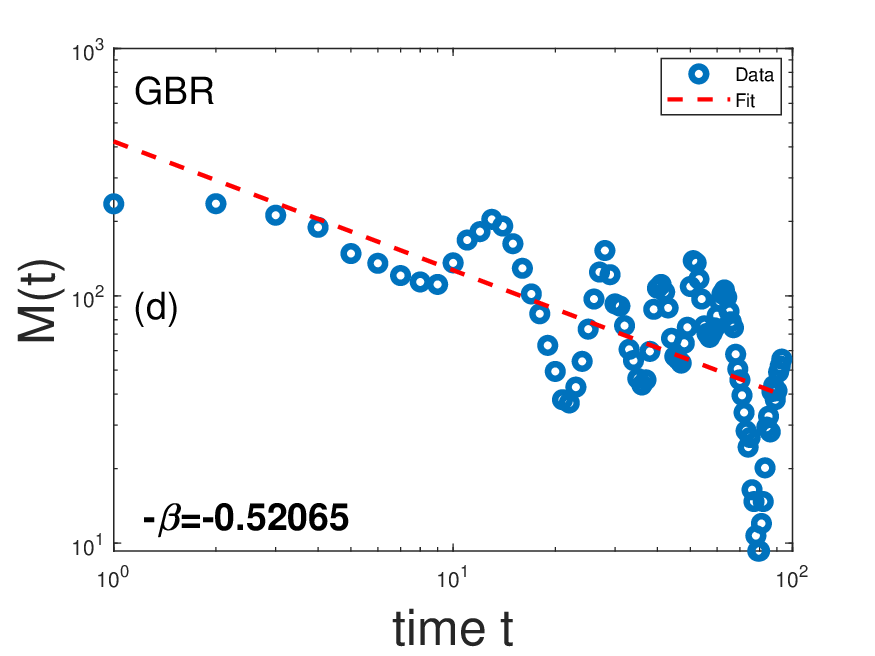}
    \includegraphics[width=4.2cm]{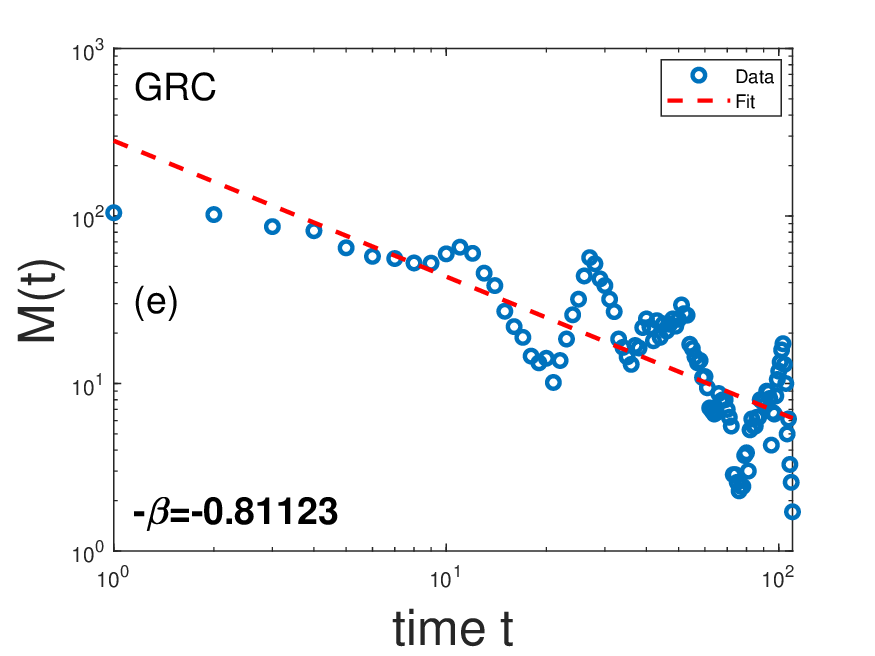}
    \includegraphics[width=4.2cm]{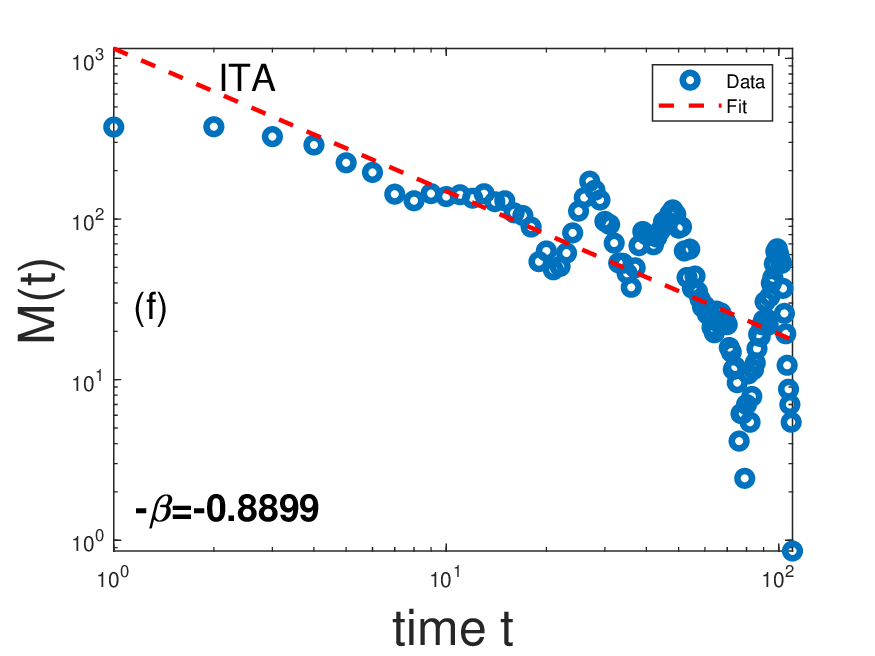}
    \includegraphics[width=4.2cm]{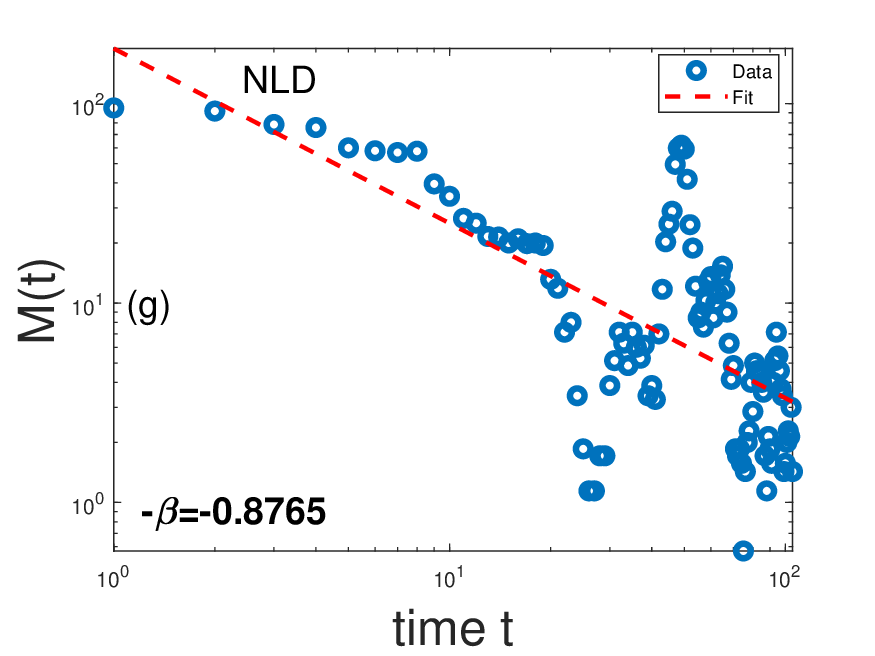}
    \includegraphics[width=4.2cm]{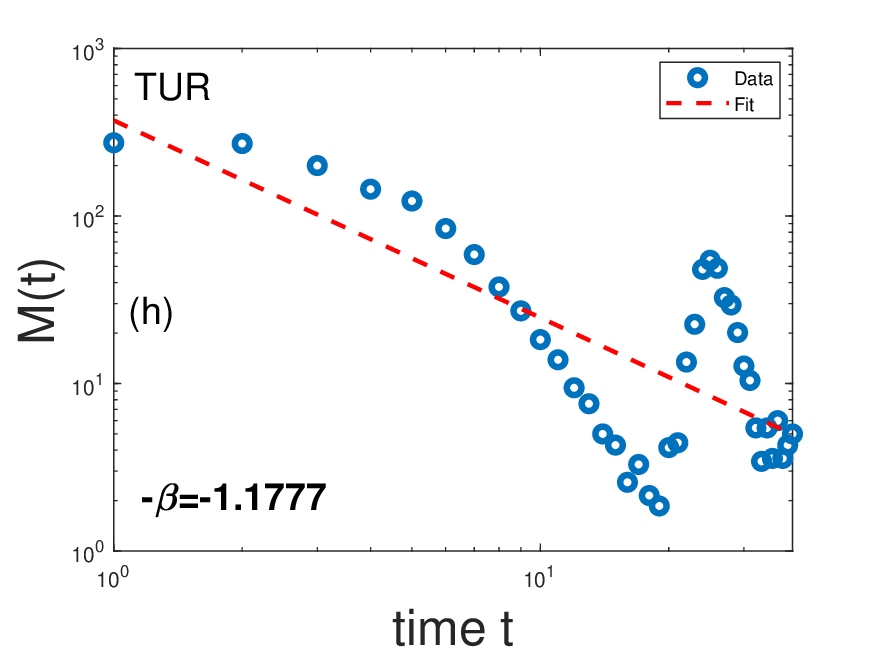}
    \caption{Time-dependence of the mortality $M(t)$ versus time $t$ for eight countries in (a) DEU, in (b) ESP, in (c) FRA, in (d) GBR, in (e) GRC, in (f) ITA, in (g) NLD, and in (h) TUR. } \label{Europe_decay}
\end{figure}
It is important to see how the deaths resulting from the Covid-19 epidemic in these countries have decreased over time recently. Because, we showed that the death distributions for this period obeyed the inverse power-law. Therefore, the time-dependent decay of this period is a sign that it may comply with a universal law. Time dependence for these countries is given in Fig.\ref{Europe_decay}. These results show that the time-dependent decay of the Covid-19 tail obeys the power-law: 
\begin{eqnarray} \label{inverse_power_decay}
	M (t) \sim t^{ - \beta}
\end{eqnarray}
where $t$ is the re-scaled time step and $\beta$ is the decay exponent of the mortality for different countries.

As can be seen from the sub-figures in Fig.\ref{Europe_decay}, $\beta$ takes values in the range $0 < \beta <2$ for different countries. For instance: for DEU $\beta=1.0347$, for ESP  $\beta=0.58675$, for FRA $\beta=0.63954$, for GBR $\beta=0.52065$, for GRC $\beta=0.81123$, for ITA $\beta=0.8899$, for NLD $\beta=0.8765$, and for TUR $\beta=1.1777$. 

\begin{figure}[!]
    \centering
    \includegraphics[width=4.2cm]{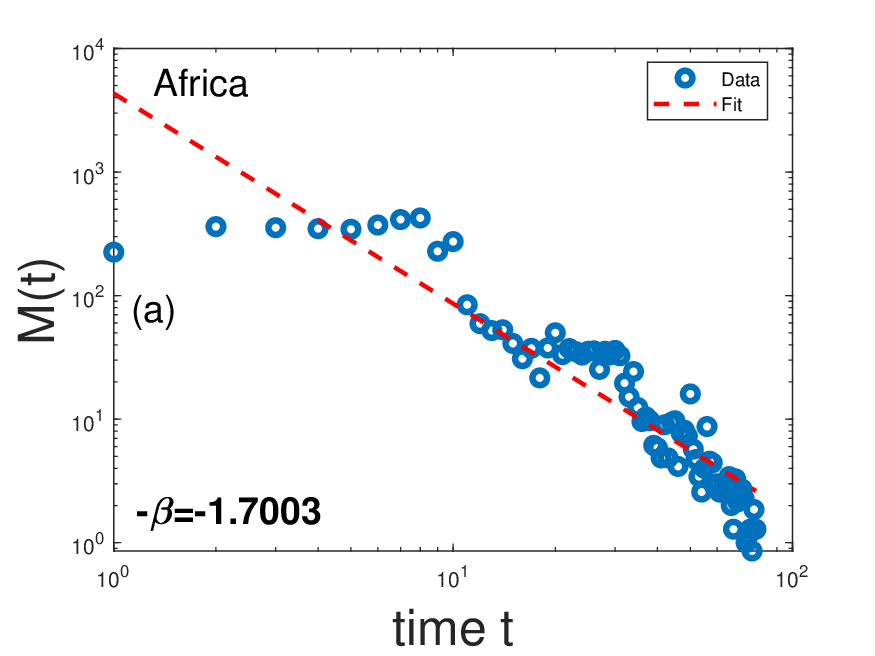}
    \includegraphics[width=4.2cm]{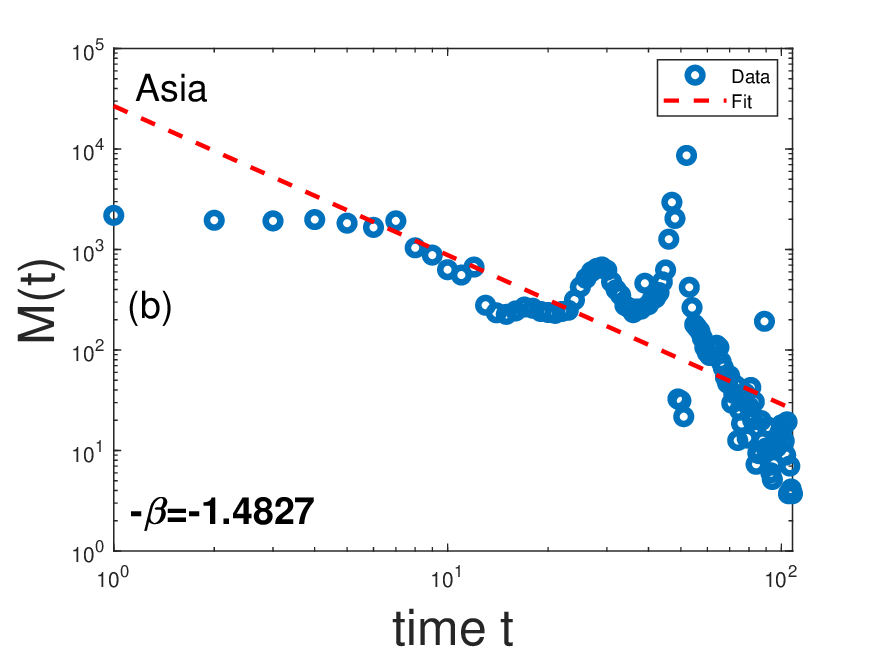}
    \includegraphics[width=4.2cm]{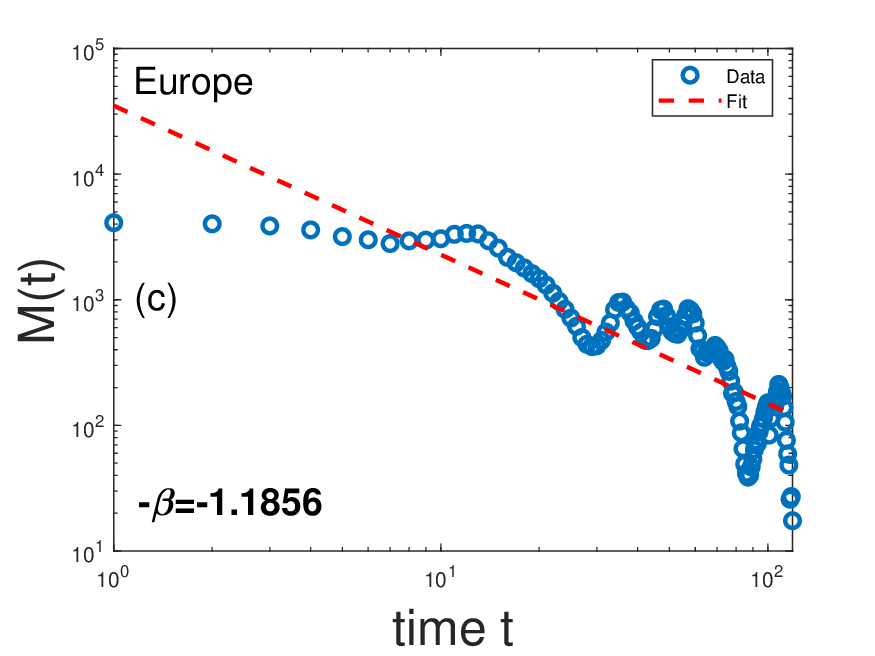}    
    \includegraphics[width=4.2cm]{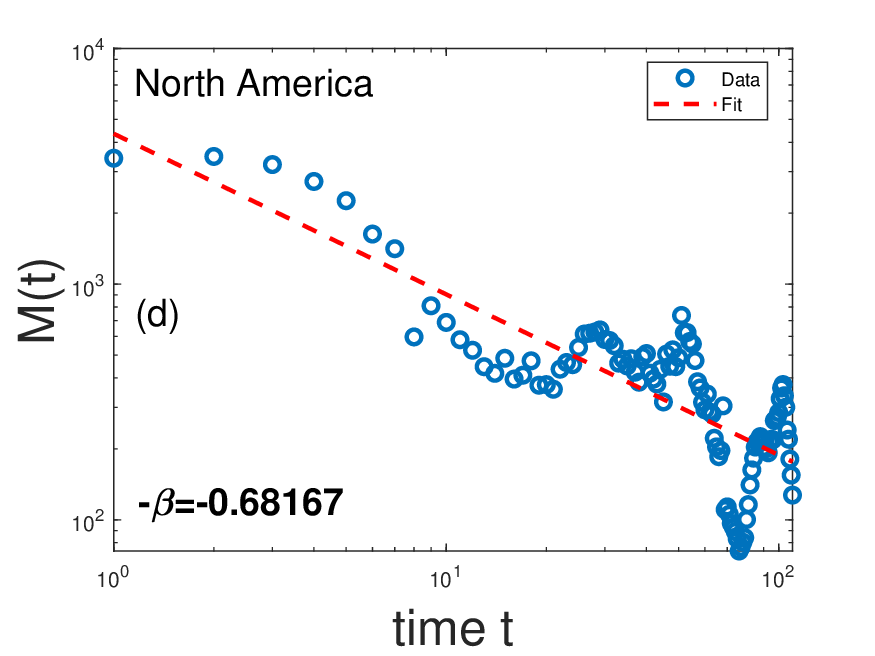}
    \includegraphics[width=4.2cm]{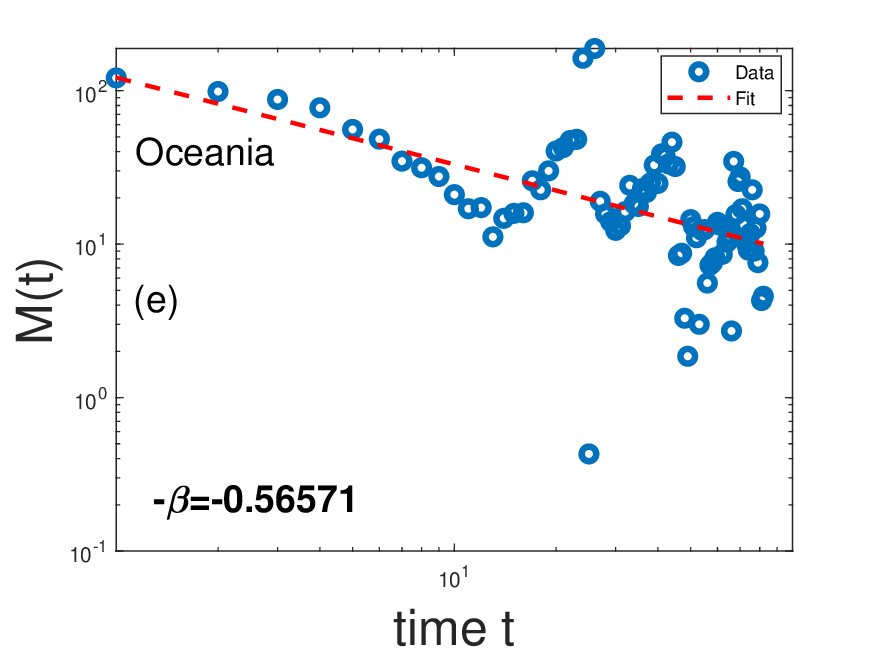}
    \includegraphics[width=4.2cm]{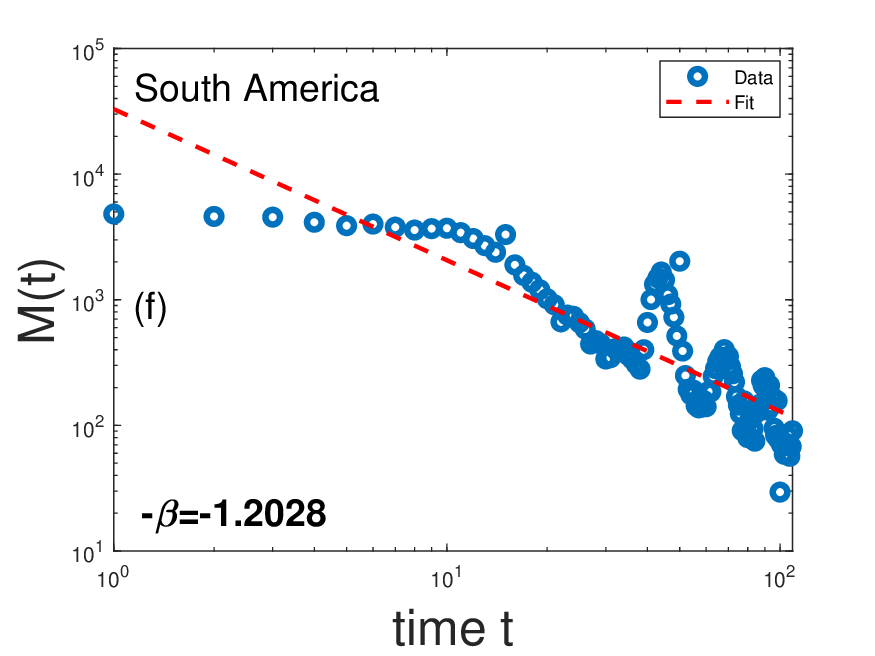}       
    \caption{Time-dependence of the mortality $M(t)$ versus time $t$ for six continentals in (a) Africa, in (b) Asia, in (c) Europe, in (d) North America, in (e) Oceania and in (f) South America.}
    \label{contin_decay}
\end{figure}

Similarly, we also analyze the time-dependent behavior of the mortality for six continents. It can be seen in Fig.\ref{contin_decay} also complies with Eq.(\ref{inverse_power_decay}). For continents, the $\beta$ value takes values between 0 and 2. For instance: for Africa $\beta=1.7003$, for Asia $\beta=1.4827$, for Europe $\beta=1.1856$, for North America $\beta=0.68167$, for Oceania $\beta=0.56571$, for South America $ \beta=1.2028$.

One can see that the time-dependent decay or relaxation in deaths in the Covid-19 tails also follows the power law like mortality distributions. As is known, relaxation processes can be modeled in terms of diffusion. If we evaluate the problem by associating it with the diffusion process, we can classify the time-dependent behavior of the Covid-19 tail as follows: 
\begin{equation} \label{type_power}
  M(t) \sim  t^{ - \beta}, \quad  \begin{cases} 
    \beta = 1 & \text{normal diffusion} \\
    0 \le \beta < 1 & \text{sub-diffusion} \\
    1<\beta \le 2 & \text{super-diffusion}
  \end{cases}
\end{equation}
We reported these novel and interesting results for Covid-19 tails for the first time in the present study.

\section{Discussion and Conclusion}

In this study, we analyzed the last tails of Covid-19 for eight countries and six continental regions to discuss the universal behavior in the distribution and time dependence of the tails. Firstly, we show that the deaths in the latest tails of Covid-19 in 2023 for the different countries and different regions we analyzed can be represented by a distribution function that obeys the inverse power-law: $P (m) \sim m^{ - \alpha}$ with $0 < \alpha <2$. Secondly, we analyzed the time dependence of the tail of Covid-19 death data for the same countries and the same continents and showed that the number of deaths decreases with an inverse power-law: $M (t) \sim t^{ - \beta}$ with $0 < \beta <2$. These novel and interesting results are reported in this work. 

It should be noted that, so far, many statistics have been applied to Covid-19 data to estimate the spreading dynamics of the pandemic. 
Weibull and modified Weibull distributions can be considered among the most widely used \cite{Moreau2020,Alahmadi2022,Liu2021,Moreau2020,Almongy2021_RP}. Similarly, different distributions were applied to the epidemic data.
For example; Kaniadakis et al \cite{Kaniadakis2020_SP} apply $\kappa$-statistics distribution to some empiric Covid-19 data from several countries. They showed that empirical data of Covid-19 peaks are consistent with $\kappa$-statistics distribution. Similarly, Tsallis and Tirnakli \cite{Tsallis2020_FP} proposed a $q$-statistical functional form to describe Covid-19 data for several countries \cite{Tsallis2020_FP}. In a recent study, Kaniadakis also showed that when power-law interactions are taken into account, the SIR model can give more harmonious results and the compartment population distributions evolve in time exhibiting power-law tails \cite{Kaniadakis2024}. On the other hand, in many studies in the literature, it has been reported that some of the growth or decay tails of the independent waves of the epidemic data between 2020 and 2023 are exponential \cite{Komarova2020_Royal,Remuzzi2020_covid,Li_2021_C,Bartolomeo2021_Infection}, sub-exponential \cite{maier2020effective} or some follow the inverse power-law decay \cite{Komarova2020_Royal,Blasius2020_Chaos,Kaniadakis2020_SP,Tsallis2020_FP,Kaniadakis2024}.  These analyses may indicate that Covid-19 data of the different countries may have different characteristics at the same periods. Unlike previous studies, in this study, we were only interested in the final Covid-19 peak tails in 2023 for various countries and continentals not independent Covid-19 peaks between 2020 and 2023 periods, and we report that the latest tails of Covid-19 data decay with power-law like form which may denotes universal character as well in the previous studies \cite{Pareto1897,Auerbach_1913,Zipf_1949,Gutenberg_1944,Estoup_1916,Neukum_1994,Price_1965,Adamic_2000,Lu_1991,Roberts_1998,Adamic_2000,Zanette_2001,Yule_1922,Albert_2002}.

If we return to our work again, to the best of our knowledge, we can state that the physical origin of the power law distribution and decay mechanisms are a controversial issue. Many alternative mechanisms have been suggested to generate power-law behavior so far. For example, it has been shown in various studies that the sum of different exponential distributions can generate a power-law distribution \cite{Miller1957,Reed2002}, and that a random walk and critical phenomena \cite{Newman_2005} or taxonomy model of biological species \cite{Yule1922,Yule1925} can generate a power-law behavior. Additionally, it is possible to add different models \cite{Carlson1999,Carlson2000,Sneppen1996CoherentNS} to them. In this work, we do not currently provide a mathematical model of how various stochastic effects produce power-laws in the tail of Covid-19 in our example, however, we can discuss possible predictions. The Covid-19 pandemic, which started in 2020 and infected approximately one billion people and caused more than seven million deaths by the end of 2023, has been thoroughly suppressed thanks to the preventive measures of governments and the developed vaccines. Despite the emergence of different variants, the virus rose to an acceptable level of herd immunity and lost its effect. The tail region, where Covid-19 begins to lose its effect, is therefore expected to differ significantly from the dynamics of the intermediate waves in the time series. Because, we can evaluate that the fact that all stochastic parameters that play a role in suppressing the lethality of Covid-19 in the tail region have been maximized may have played a role in generating the power-law dynamics.

As a result, as known that power-law dynamics is an active research topic in many fields of science such as physics, computer science, linguistics, geophysics, neuroscience, sociology, economy and more. Therefore, it is very important efforts to investigate the origins of power-law relations and to observe and verify them in physics and other areas. In this study, we have shown that the power-law dynamic can also take place in a pandemic.



\bibliography{Covid_power}


\end{document}